\documentclass{article}
\usepackage[preprint]{spconf}
\usepackage{amsmath,graphicx,url}

\title{BANDING VS. QUALITY: PERCEPTUAL IMPACT AND OBJECTIVE ASSESSMENT}
\name{Luk\'{a}\v{s} Krasula, Zhi Li, Christos G. Bampis, Mariana Afonso, Nil Fons Miret, Joel Sole }
\address{Netflix Inc., Los Gatos, CA, USA, 95032. \{lkrasula, zli, christosb, mafonso, nfonsmiret, jsole\}@netflix.com}

\begin{document}
%
\maketitle
\begingroup\renewcommand\thefootnote{\copyright}
\footnotetext{IEEE 2022. Personal use of this material is permitted. Permission from IEEE must be obtained for all other uses, in any current or future media, including reprinting/republishing this material for advertising or promotional purposes, creating new collective works, for resale or redistribution to servers or lists, or reuse of any copyrighted component of this work in other works.}
\endgroup
\begin{abstract}
Staircase-like contours introduced to a video by quantization in flat areas, commonly known as banding, have been a long-standing problem in both video processing and quality assessment communities. The fact that even a relatively small change of the original pixel values can result in a strong impact on perceived quality makes banding especially difficult to be detected by objective quality metrics. In this paper, we study how banding annoyance compares to more commonly studied scaling and compression artifacts with respect to the overall perceptual quality. We further propose a simple combination of VMAF and the recently developed banding index, CAMBI, into a banding-aware video quality metric showing improved correlation with overall perceived quality.
\end{abstract}
\section{Introduction}
\label{sec:intro}

Recent years have brought significant improvements in capturing, processing, and displaying images and videos. This goes hand in hand with increased expectations of the observers regarding picture quality. Objective quality metrics have played a crucial role in optimization of video processing pipelines and ensuring high quality of viewing experience. Despite the relatively high reliability of metrics such as SSIM \cite{wang2004image} or VMAF \cite{li2016toward}, especially when it comes to detecting standard compression artifacts (e.g. blocking, blurring, etc.), there are still a few scenarios where these metrics perform significantly worse. One such example is banding.

Banding is a common name for staircase-like contours appearing in low frequency regions of video frames. They can be caused by multiple factors, one of the most prominent being the quantization step in a video codec. Even a relatively small change of the pixel values can produce severe banding which makes it very hard to be detected by a general purpose objective quality metrics \cite{cambi_pcs21}. A number of specialized banding detectors have been proposed in the literature. They typically approach the problem as false edge detection using local statistics in the frame \cite{lee2006two}, \cite{huang2016understanding}, or as false segment identification in either block-wise \cite{jin2011composite}, \cite{wang2014multi} or pixel-wise manner \cite{wang2016perceptual}, \cite{baugh2014advanced}. More recent banding estimators exploit the advances in machine learning, either by training a set of human visual system (HVS) inspired hyperparameters \cite{bband} or a deep neural network \cite{dbi_icassp2021}. 

In our previous work, we have developed a Contrast Aware Multi-scale Banding Index (CAMBI) \cite{cambi_pcs21}. CAMBI directly looks for steps in areas with high probability of banding occurrence and uses properties of HVS to determine if and how much they will be visible to human observers.

Even though some subjective tests targeting banding were conducted, e.g. \cite{wang2016perceptual} or \cite{dbi_icassp2021}, they were primarily designed for training and testing of banding detectors and did not provide insights into subjective perception of the artifacts themselves. 

In this paper, we evaluate the perceptual impact of banding on perceived quality compared to other more typical and better understood video compression artifacts. This is achieved by a large-scale subjective study including videos from \cite{cambi_pcs21}, for which banding annoyance scores are available, mixed with videos from the Netflix dataset used for development of VMAF 4K model \cite{li2018vmafjourney}.

Furthermore, we show that even a simple linear combination of VMAF and CAMBI is able to sufficiently fit the study data. We refer to this combination as VMAF\textsubscript{BA} -- a banding-aware VMAF. The performance of VMAF\textsubscript{BA} on several datasets reveals that considering banding can lead to improvement in correlation with subjective quality scores not only on our banding-specific database but in case of general video compression as well.

The paper is organized as follows: Section \ref{sec:subjexp} provides details on the subjective experiment and quantification of the impact of banding on the overall perceived quality in the context of video compression. Objective metrics behavior with respect to the obtained subjective data, as well as the derivation of the banding-aware combination of metrics are discussed in Section \ref{sec:objexp}. Section \ref{sec:combi} shows the effectiveness of the proposed combination on multiple video datasets and, finally, Section \ref{sec:conclusion} concludes the paper and discusses future work.

\section{Subjective Experiment}
\label{sec:subjexp}

\begin{figure}[tb]
\centering
\begin{minipage}[b]{0.48\linewidth}
  \centering
  \centerline{\includegraphics[width=4.5cm]{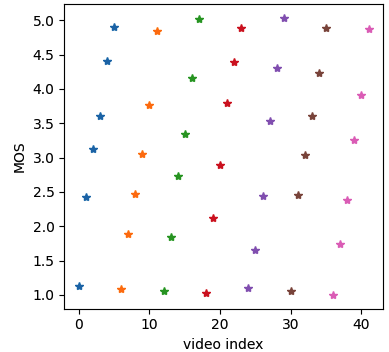}}
\end{minipage}
\caption{Distribution of the original MOS scores for the videos from VMAF 4K dataset. Each color represents a different source content.}
\label{fig:vmaf_4k}
\end{figure}

The goal of the study was to compare the annoyance of banding to the common compression artifacts. In other words, we want to place the videos affected by banding on a similar quality scale as the videos used to train VMAF. To achieve that, we selected 42 videos (6 distorted videos coming from 7 sources) from the VMAF 4K dataset \cite{li2018vmafjourney}. These were videos with encoding resolution from 2160p to 216p compressed by H.264 encoder. We made sure that the selected sequences span the entire quality scale from "Bad" to "Excellent". The distribution of the original mean opinion scores (MOS) for these videos is shown in Fig. \ref{fig:vmaf_4k}. It can be seen that all quality levels are well represented. Note that the scores are on a standard Absolute Category Rating (ACR) Scale from ITU-T Rec. P.910 \cite{recommendationitu}.

The second half of the sequences (42 videos, 6 distorted videos from 7 sources) was chosen from the dataset used for validation of CAMBI \cite{cambi_pcs21}. Here, the encoding resolutions are 2160p, 1440p, or 1080p, the used codec is libaom, and banding is the dominant distortion. All selected videos have 8 bits per color channel.

\subsection{Subjective methodology and observers}
We ran the subjective test using the ACR methodology with a continuous quality scale from 0 to 100, annotated with 5 labels (0 - Bad, 25 - Poor, 50 - Fair, 75 - Good, 100 - Excellent). This modification is allowed in the ITU-T Rec. P.910 \cite{recommendationitu} and provides more freedom to the observers to express their opinion and is generally preferred by the subjects over the discrete scale \cite{Belz_discretevs.}. Each test session took approximately 25 minutes in total.

A total of 42 observers were recruited from Netflix employees who do not work directly in video encoding or quality assessment. All observers evaluated all videos. The study can be considered "semi-controlled" because despite being remote, the subjects were asked to set up the viewing conditions according to specific instructions -- dim ambient light, 75\% screen brightness, and a normalized viewing distance of 1.5 times the physical screen height. 

Before the beginning of the test, we conducted a training session with 6 videos created from sources excluded from the main test. 3 of the videos had different level of standard compression artifacts while the other 3 had different severity of banding to give the participants an idea of what distortions to expect. Given the volunteering basis of the study, the subject reliability was expected to be better than in a typical crowdsourcing study. This is confirmed by the data in the next section.

\subsection{Results processing}
We used the advanced data analysis technique for tests under challenging conditions from Annex E of ITU-T Rec. P.910 to process the results and obtain the MOS. Firstly, we perform a data reliability check by comparing the obtained results for the videos from VMAF 4K dataset with their original scores from the previous experiment. Then we analyze the \textit{overall perceptual quality} of videos with banding and compare them to their original \textit{banding annoyance} scores.

\subsubsection{Reliability check on videos from VMAF 4K dataset}
\begin{figure}[tb]
\begin{minipage}[b]{0.48\linewidth}
  \centering
  \centerline{\includegraphics[width=4.0cm]{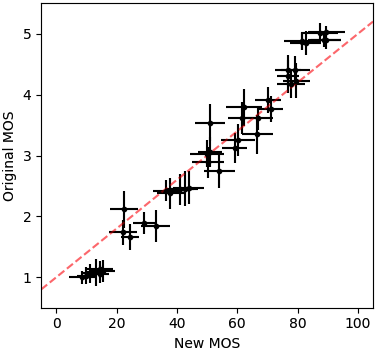}}
  \centerline{(a) VMAF 4K dataset videos}\medskip
\end{minipage}
\hfill
\begin{minipage}[b]{0.48\linewidth}
  \centering
  \centerline{\includegraphics[width=4.0cm]{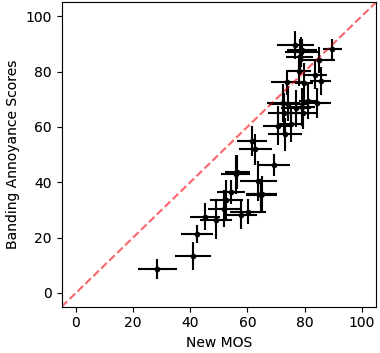}}
  \centerline{(b) Banding dataset videos}\medskip
\end{minipage}
\caption{Comparison of the original vs. new MOS for both parts of the dataset with their respective confidence intervals. The red dashed line is connecting the lowest and the highest point of each scale.}
\label{fig:subj_res}
\end{figure}

The scatter plot of the original vs. new MOS can be found in Fig. \ref{fig:subj_res}(a). All points are accompanied with their respective confidence intervals. The scores seem to reasonably follow the $45^{\circ}$ line and also reach very high Pearson Linear Correlation Coefficient (PLCC) and Spearman Rank-Order Correlation Coefficient (SROCC) -- \textbf{PLCC = 0.987} and \textbf{SROCC = 0.989}, respectively. These numbers are similar to the typical correlation found when discrete and continuous scales are compared \cite{Belz_discretevs.}.

More importantly, there are no cases where the rank-order of any two videos is flipped while the scores are statistically significantly different in both experiments. In other words, every time a pair of videos has different rank-order, the confidence intervals are overlapping in at least one of the experiments.

The size of the confidence intervals is also comparable which suggests that the higher number of observers in the semi-controlled study (42 vs. 24) was able to compensate for the limited control over the viewing environment, while not requiring hundreds of observers and thorough reliability checks as in a regular crowdsourcing study \cite{Hossfeld2014_crowdsourcing}. 

We can notice a small scale compression effect in the new experiment, i.e. the new MOS do not reach as high and as low values as their original counterparts. This is most likely connected to the use of continuous instead of the discrete scale where some degree of saturation at the ends is expected. Overall, the data seems to be very well aligned and thus can be considered reliable.

\subsubsection{Comparison of banding annoyance and overall quality}
The scores for videos from the banding dataset are plotted against their \textit{banding annoyance} scores in Fig. \ref{fig:subj_res}(b). The first observation is that most of the points are distributed below the $45^{\circ}$ line which can be interpreted as the new MOS being generally higher than \textit{banding annoyance}. This is consistent with our expectations as even the most severe banding in the dataset does not distort the video as much as the strongest compression setting present in the set.

Arguably the most interesting observation is a very high level of linearity between the two scales -- \textbf{PLCC = 0.916} and \textbf{SROCC = 0.894}, respectively -- despite the difference in the task. This suggests that an objective banding detector able to reliably predict the annoyance of banding could potentially be used to aid a generic quality metric to become banding-aware. 

A careful reader can also notice a slight increase in the size of the confidence intervals in spite of the higher number of observers (42 vs. 23 in the banding dataset). This is likely caused by lower inter-observer agreement coming from the higher difficulty of the task -- it is generally more difficult to compare different types of artifacts than to evaluate a single distortion.

The scores obtained for videos from both parts of the dataset will be used together in the next section to test the abilities of objective metrics.

\section{Objective Evaluation of the Dataset}
\label{sec:objexp}

As discussed in the Introduction, as well as in \cite{cambi_pcs21}, generic objective quality metrics struggle with evaluating videos with banding. Table \ref{tab:obj_dataset} shows the performance of established image and video quality metrics -- namely PSNR, SSIM \cite{wang2004image}, MS-SSIM \cite{wang_mssim}, VMAF \cite{li2016toward}, and CAMBI \cite{cambi_pcs21} -- on the dataset described in this paper.

The performance is measured using PLCC, SROCC, and AUC\_BW \cite{lukas_c0}. Unlike the standard correlation coefficients, AUC\_BW takes into account uncertainty of the subjective scores and tests metrics' ability to correctly identify the higher quality video from any pair with statistically significantly different scores. AUC\_BW of 0.5 corresponds to a random guessing (equivalent to a correlation of 0) while the value of 1 suggests a perfect performance. 

\begin{table}[htb]
\centering
\small
\caption{Performance of objective metrics on the new dataset.}
\label{tab:obj_dataset}
\begin{tabular}{|l|c|c|c|}
\hline
 & PLCC & SROCC & AUC\_BW \\
 \hline
PSNR & 0.500 & 0.384 & 0.733 \\
SSIM & 0.677 & 0.552 & 0.854 \\
MS-SSIM & 0.622 & 0.404 & 0.787 \\
VMAF & \textbf{0.842} & \textbf{0.677} & \textbf{0.912} \\
CAMBI & 0.102 & 0.141 & 0.511 \\
\hline
\end{tabular}
\end{table}

After an initial look, the results do not significantly deviate from our expectations. VMAF outperforms the other metrics while CAMBI fares poorly as it is only sensitive to banding artifacts. On the other hand, it is unusual for SROCC values to be that much lower than PLCC, especially when no mapping had been applied before calculating the PLCC. This suggests we should take a closer look at the results.

\begin{figure}[tb]
\begin{minipage}[b]{0.48\linewidth}
  \centering
  \centerline{\includegraphics[width=4.0cm]{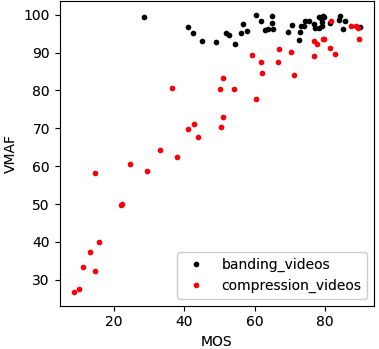}}
  \centerline{(a) VMAF vs. MOS}\medskip
\end{minipage}
\hfill
\begin{minipage}[b]{0.48\linewidth}
  \centering
  \centerline{\includegraphics[width=4.0cm]{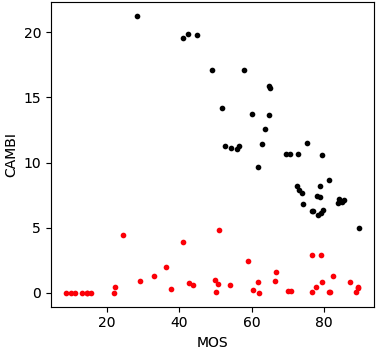}}
  \centerline{(b) CAMBI vs. MOS}\medskip
\end{minipage}
\caption{Scatter plot of objective and subjective scores. The banding and compression parts of the dataset are depicted in different colors for illustration purposes.}
\label{fig:obj_res}
\end{figure}

Fig. \ref{fig:obj_res}(a) shows VMAF scores plotted against MOS. For better illustration, the two types of videos in the dataset are depicted in different colors. VMAF's inability to correctly evaluate quality of the videos with banding is obvious from the plot. When we look at the results for CAMBI in Fig. \ref{fig:obj_res}(b) the situation looks almost exactly opposite. The two plots suggest that a \textit{banding-aware VMAF} (VMAF\textsubscript{BA}) could potentially be obtained by a simple linear combination of the two metrics. 

\begin{figure}
  \centering
\begin{minipage}[b]{0.48\linewidth}
  \centering
  \centerline{\includegraphics[width=4.5cm]{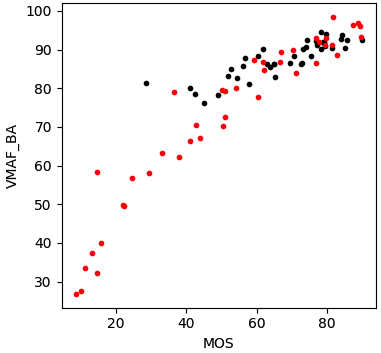}}
\end{minipage}
    \caption{Scatter plot of VMAF\textsubscript{BA} vs. MOS. The banding and compression parts of the dataset are depicted in different colors for illustration purposes.}
    \label{fig:vmaf_ba}
\end{figure}

In order to maintain VMAF\textsubscript{BA} equal to 100 when comparing a reference video with no banding to itself, we explored the combination in the form of
\begin{equation}
    \mbox{VMAF\textsubscript{BA}} = \mbox{VMAF} - \alpha \times \mbox{CAMBI}.
\end{equation}
By running an optimization procedure maximizing SROCC for the above described dataset, we identified $\alpha = 0.85$ as a reasonable solution. To avoid negative values, we clip the scale at 0. The scatter plot for this combination is depicted in Fig. \ref{fig:vmaf_ba}. 

We can see that the ability to correctly assess the quality of videos with banding significantly improved which is also reflected in the performance measures -- \textbf{PLCC = 0.916}, \textbf{SROCC = 0.937}, and \textbf{AUC\_BW = 0.993}.

\begin{table*}[htb]
\centering
\small
\caption{Performance of objective metrics on 7 datasets.}
\label{tab:plcc_srocc}
\begin{tabular}{|l|c|c|c|c|c|c|c|}
\hline
\textbf{PLCC} & VMAF4K \cite{li2018vmafjourney} & NFLX \cite{li2016toward} & VMAF+ \cite{vmaf_plus} & VQEGHD3 \cite{vqeghd3} & LIVEvideo \cite{live_video} & LIVEmobile \cite{live_mobile} & CSIQVQA \cite{csiqvqa} \\
\hline
VMAF\textsubscript{BA} & \textbf{0.899} & \textbf{0.944} & \textbf{0.906} & \textbf{0.946} & 0.700 & 0.889 & 0.612 \\
VMAF & 0.890 & 0.937 & 0.902 & 0.936 & \textbf{0.709} & \textbf{0.893} & 0.608 \\
SSIM & 0.708 & 0.750 & 0.734 & 0.879 & 0.630 & 0.717 & 0.712 \\
MS-SSIM & 0.605 & 0.729 & 0.693 & 0.871 & 0.626 & 0.711 & \textbf{0.738} \\
\hline
\end{tabular}
\begin{tabular}{|l|c|c|c|c|c|c|c|}
\hline
\textbf{SROCC} & VMAF4K \cite{li2018vmafjourney} & NFLX \cite{li2016toward} & VMAF+ \cite{vmaf_plus} & VQEGHD3 \cite{vqeghd3} & LIVEvideo \cite{live_video} & LIVEmobile \cite{live_mobile} & CSIQVQA \cite{csiqvqa} \\
\hline
VMAF\textsubscript{BA} & \textbf{0.899} & \textbf{0.926} & \textbf{0.904} & \textbf{0.939} & 0.719 & 0.861 & 0.622 \\
VMAF & 0.893 & 0.922 & 0.901 & 0.924 & \textbf{0.726} & \textbf{0.863} & 0.615 \\
SSIM & 0.751 & 0.806 & 0.722 & 0.904 & 0.685 & 0.709 & 0.698 \\
MS-SSIM & 0.625 & 0.765 & 0.679 & 0.895 & 0.692 & 0.699 & \textbf{0.749} \\
\hline
\end{tabular}
\end{table*}

One potential shortcoming of the presented dataset, obvious from the Fig. \ref{fig:obj_res}, is the absence of sequences with heavy banding and compression artifacts together, i.e. low VMAF and high CAMBI. To get an idea about the robustness of VMAF\textsubscript{BA}, we test it on a number of available datasets. 

\section{General Performance of the Proposed Combination}
\label{sec:combi}

The PLCC and SROCC values for VMAF\textsubscript{BA}, VMAF, SSIM, and MS-SSIM on seven datasets (3 private and 4 public) are presented in Table \ref{tab:plcc_srocc}. Note that AUC\_BW values are not shown as not all of the datasets provide the confidence intervals with their scores, nevertheless, they follow the same trends when available. 

The main purpose is to determine whether VMAF's performance does not get negatively impacted by the proposed modification into a banding-aware metric. When we study the Table \ref{tab:plcc_srocc}, the correlation drops only in two out of seven tested cases. Moreover, the decreased values still remain significantly higher than for SSIM and MS-SSIM. We can also see a slight improvement on the other five databases. 

Overall, there seems to be a performance benefit of combining VMAF and CAMBI even in this simple and easily interpretable way. Nevertheless, VMAF\textsubscript{BA} also shares the shortcomings of VMAF, as evidenced by the performance on CSIQVQA database \cite{csiqvqa} where some videos with artifacts unrelated to compression are present. The result remains worse than for other metrics, despite the slight improvement.

Even though the presented results suggest good robustness of VMAF\textsubscript{BA}, we are planning to further investigate the specific cases with low VMAF and high CAMBI and the consecutive strong interaction between banding and other artifacts as part of our future work.

\section{Conclusion and Future Work}
\label{sec:conclusion}

In this paper, we investigated the impact of banding on the overall perceptual quality compared to common scaling and compression (H.264) artifacts. Using a large-scale subjective study, we showed the relationship between banding annoyance and overall quality to be fairly linear.

Our collected data confirmed the inaccuracy of objective metrics when evaluating banding, as well as the inability of CAMBI to capture other types of artifacts. Furthermore, we formulated a simple combination of VMAF and CAMBI into VMAF\textsubscript{BA} -- a banding-aware, generic video quality metric. We demonstrated that the proposed extension does not significantly decrease VMAF's correlation with subjective scores on datasets with generic compression distortions and can even lead to slight improvements overall.

To continue this work, we plan to further investigate the interaction between heavy compression and banding both perceptually and from the perspective of objective metrics. We will also investigate potential benefits of incorporating CAMBI directly into VMAF as one of its elementary features. 

Implementations for both VMAF and CAMBI are available at \url{https://github.com/Netflix/vmaf}.


\bibliographystyle{IEEEbib}
\bibliography{strings,refs}

\end{document}